\documentclass[letterpaper,12pt]{article}
\usepackage[utf8]{inputenc}
\usepackage[T1]{fontenc}
\usepackage[margin=1in]{geometry}
\usepackage{helvet}
\usepackage{graphicx}
\usepackage{wrapfig}
\usepackage[english]{babel}
\usepackage[square,numbers,sort&compress]{natbib}
\usepackage{hyperref}
\hypersetup{colorlinks=true, urlcolor=blue, citecolor=blue, linkcolor=blue, bookmarksopen=true, pdftoolbar=true, pdfmenubar=true, bookmarksopenlevel=\maxdimen, bookmarksnumbered=true, bookmarkstype=toc, debug=true}
\usepackage[compact]{titlesec}
\usepackage{authblk}
\usepackage{caption}
\newcommand\tab[1][1cm]{\hspace*{#1}}
\title{ \textbf{The Persistent Mystery of Collisionless Shocks} \\ }
\author[1]{Katherine Goodrich}
\author[2,3]{Steven Schwartz}
\author[4]{Lynn Wilson III}
\author[5]{Ian Cohen}
\author[5]{Drew Turner}
\author[6]{Amir Caspi}
\author[6]{Keith Smith}
\author[6]{Randall Rose}
\author[7]{Phyllis Whittlesey}
\author[8]{Ferdinand Plaschke}
\author[9]{Jasper Halekas}
\author[9]{George Hospodarsky}
\author[6]{James Burch}
\author[10]{Imogen Gingell}
\author[4]{Li-Jen Chen}
\author[11]{Alessandro Retino}
\author[12]{Yuri Khotyaintsev}

\affil[1]{West Virginia University, Morgantown, WV}
\affil[2]{Laboratory for Atmospheric and Space Physics, University of Colorado, Boulder CO}
\affil[3]{also Emeritus Professor, Imperial College London, London, UK}
\affil[4]{Goddard Space Flight Center, Laurel, MD}
\affil[5]{Applied Physics Laboratory, John Hopkins University, Laurel, MD}
\affil[6]{Southwest Research Institute, San Antonio, TX}
\affil[7]{Space Sciences Laboratory, University of California, Berkeley, CA}
\affil[8]{Technische Universit{\"a}t Braunschweig, Branschweig, Germany}
\affil[9]{University of Iowa, Iowa City, IA}
\affil[10]{University of Southhampton, Southhampton, England}
\affil[11]{Laboratoire de Physique des Plasmas, Paris, France}
\affil[12]{Swedish Institute of Space Physics, Uppsala, Sweden}

\affil[13]{University of California, Los Angeles, CA}
\affil[14]{Princeton University, Princeton, NJ}
\affil[15]{University of Hawaii, Honolulu, HI}
\affil[16]{University of Arizona, Tucson, AZ}
\affil[17]{University of Maryland, College Park, MD}
\affil[18]{University of Alaska, Fairbanks, AK}
\affil[19]{Smithsonian Astrophysical Observatory, Cambridge, MA}
\affil[20]{University of Chicago, Chicago, IL}
\affil[21]{National Institute of Astrophysics, Rome, Italy}
\affil[22]{University of Toulouse, Toulouse, France}
\affil[23]{University of New Hampshire, Durham, NH}
\affil[24]{Queen Mary University, London, UK}
\affil[25]{University of Versailles - Paris-Saclay University, Paris, France}
\begin{document}
\begin{titlepage}
\maketitle

{\small \hspace{-20pt}\textbf{Co-signers:} C.T. Russell$^{13}$, 
J. TenBarge$^{14}$,
D. Malaspina$^{2}$,
C. Haggerty$^{15}$,
T. Liu$^{13}$,
D. Sibeck$^{4}$,
K. Klein$^{16}$,
I. Vasko$^{7}$,
S. Wang$^{17}$,
E. Scime$^1$,
H. Zhang$^{18}$, 
J. Juno$^{14}$, 
M. Stevens$^{19}$,
D. Caprioli$^{20}$, 
M.F. Marcucci$^{21}$, 
C. Mazelle$^{22}$, 
G. Howes$^{9}$, 
I. Gingell$^{10}$,
H. Kucharek$^{23}$, 
J. Giacalone$^{16}$,
D. Burgess$^{24}$,
B. Lemb\'{e}ge$^{25}$}\\

\thispagestyle{empty}
\end{titlepage}

\bibliographystyle{bibliography_style/apjshort}
\titlespacing{\wrapfigure}{0pt}{0pt plus 1pt minus 1pt}{0pt plus 1pt minus 1pt}

\pagebreak
\section*{Synopsis}
Collisionless shock waves are one of the main forms of energy conversion in space plasmas. They can directly or indirectly drive other universal plasma processes such as magnetic reconnection, turbulence, particle acceleration and wave phenomena. Collisionless shocks employ a myriad of kinetic plasma mechanisms to convert the kinetic energy of supersonic flows in space to other forms of energy (e.g., thermal plasma, energetic particles, or Poynting flux) in order for the flow to pass an immovable obstacle. The partitioning of energy downstream of collisionless shocks is not well understood, nor are the processes which perform energy conversion. While we, as the heliophysics community, have collected an abundance of observations of the terrestrial bow shock,  instrument and mission-level limitations have made it impossible to quantify this partition, to establish the physics within the shock layer responsible for it, and to understand its dependence on upstream conditions. This paper stresses the need for the first ever spacecraft mission specifically designed and dedicated to the observation of both the terrestrial bow shock as well as Interplanetary shocks in the solar wind. 

\pagenumbering{arabic}
\setcounter{page}{1}

\section*{Outstanding Science Questions}
Understanding shocks is vital to the understanding of our universe, from the heating and deflection of bulk flows to the acceleration of cosmic rays. Moreover, collisionless shocks directly influence our own terrestrial space environment, e.g., solar wind-magnetosphere interactions.\\
\tab Vital questions regarding collisionless shocks remain unanswered:
\begin{enumerate}
    \item What is the partition of energy across collisionless shocks?
    \item What are the processes governing energy conversion at and within collisionless shocks?
    \item How and why do these processes vary with macroscopic shock parameters?
\end{enumerate}
\tab In this paper, we will discuss why addressing these questions is of critical importance. We will also discuss how our current observational limitations prevent the heliophysics community from doing so. Additionally, we provide a mission white paper "Multi-point Assessment of the Kinematics of Shocks - MAKOS" by Goodrich et al., that identifies a path forward to make significant progress towards achieving a necessary understanding to address these questions through the implementation of MAKOS (Multi-point Assessment of the Kinematics of Shocks).\\
\vspace{-15pt}\section*{Background and Motivation}

Shocks are spatial discontinuities that form when a supersonic flow encounters an obstacle. If the medium travels faster than the speed of communication, the medium has no time to smoothly adjust its trajectory. A shock forms ahead of the obstacle  and slows the supersonic flow to subsonic speeds in order for the medium to move past the obstacle. In high density media, the shock structure and evolution are governed by particle collisions.

Shocks act as a universal energy conversion mechanism in space plasmas. There is currently no known equation of state for collisionless shocks. Such an equation of state, if it could be found, would predict how the internal energy would be reconfigured as the plasma passes through a shock in response to the deceleration, deflection, heating and compression demanded by the macroscopic shock initiation.

The most relevant collisionless shock to humans, and the one most often measured in situ, is the terrestrial bow shock. The terrestrial bow shock is also significantly more straightforward to observe relative to Interplanetary (IP) shocks in the solar wind, as it remains in the same spatial position relative to Earth (to within a few Earth radii). Therefore, we derive the majority of our knowledge of collisionless shock dynamics from the terrestrial bow shock. 

The solar wind inputs primarily bulk proton ram energy upstream of the bow shock. However, the shock outputs energy in several different forms. These include, but are not limited to, electron, proton and heavy ion  acceleration and heating, together with Poynting flux and turbulent fluctuations. Previous missions together with numerical simulations  have provided invaluable insight to the overall structure and behavior of the terrestrial bow shock, as the next section discusses (e.g., \cite{Burgess2015}). However, we will show that in order to observe the shock’s detailed fundamental behavior, we require observations specifically designed to observe the terrestrial bow shock as a primary region of interest.

\section*{Current Knowledge of Collisionless Shocks}
\begin{wrapfigure}{l}{0.5\textwidth}
\vspace{-15pt}
  \centering
    \includegraphics[width=0.48\textwidth]{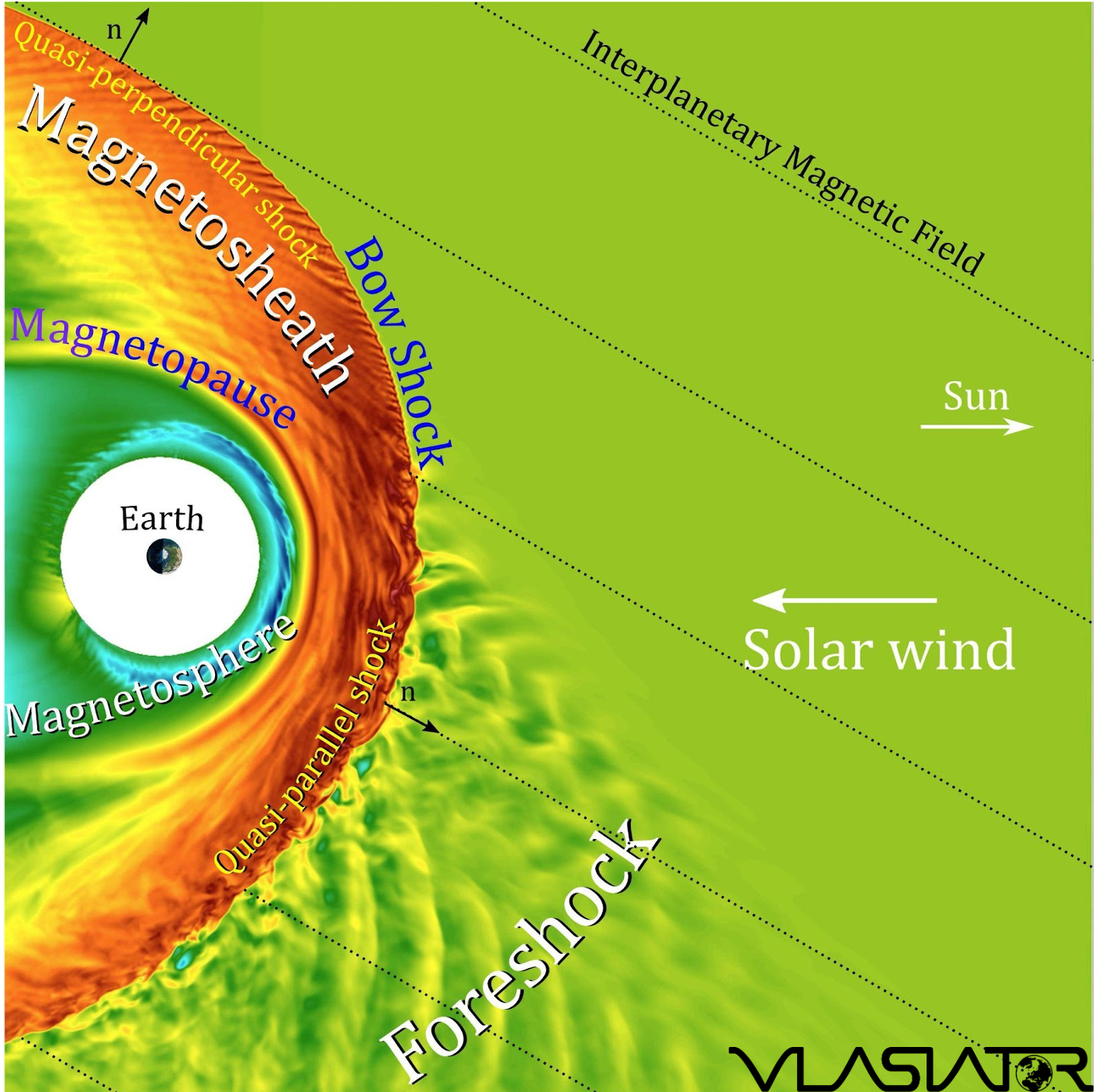}
  \caption{Global Vlasov Simulation of the terrestrial bow shock. Note the extended turbulent structure at the quasi-parallel shock (toward the bottom) by comparison to that at the quasi-perpendicular shock (toward the top).}\label{fig:Vlas}
\end{wrapfigure}

Past missions that have observed the terrestrial bow shock include MMS, THEMIS, Cluster, Wind, AMPTE, and ISEE. They have confirmed that the shock can exist as a nonstationary discontinuity. It can act as a “breathing barrier” between the solar wind and the terrestrial magnetosphere, changing in response to varying upstream conditions.

The spatial scale, energy conversion processes, and output of the shock are most heavily dependent on the orientation of the Interplanetary Magnetic Field (IMF) relative to the shock normal vector ($\hat{n}$) and the fast magnetosonic Mach number ($M_f$).  Shocks are generally categorized as either Quasi-perpendicular ($Q_{\perp}$) or Quasi-parallel ($Q_{\parallel}$) depending on whether the angle between the IMF and shock normal($\theta_{Bn}$) is greater than or less than 45$^{\circ}$.  The terrestrial bow shock also tends to grow more turbulent in nature as the Mach number ($M_f$) increases and as the plasma $\beta$ decreases.

Figure \ref{fig:Vlas} summarizes the complexity of the global structure of the terrestrial bow shock. At $Q_{\perp}$ shocks (i.e., toward the top of the figure), particle motion in the shock-normal direction is restricted by the Lorentz force to within one gyroradius in the upstream direction. Thus, $Q_{\perp}$ shocks tend to have short coherent transition regions, with quasi-static magnetic and electric fields making the largest contributions to the bulk particle dynamics. 

$Q_{\parallel}$ shocks (bottom of Figure \ref{fig:Vlas}) permit particle traversals in both directions across the shock, including well into the upstream region. Such shocks exhibit an extended transition region and are dominated by strongly varying particle  sub-populations, particle reflection with corresponding kinetic instabilities and turbulence, and particle acceleration. They can also be populated with foreshock transient events such as hot flow anomalies \cite{Schwartz2018IonObservations} and foreshock bubbles \cite{Turner2013FirstHFAs}, which can locally generate their own shocks and foreshocks \cite{Wilson2016RelativisticShock, WilsonIII2013ShockletsForeshock} in poorly understood ways. 

\section*{Outstanding Questions and Necessary Measurements}
While it is known that collisionless shocks perform energy conversion, specifically to process the bulk flow kinetic energy density \cite{WilsonIII2014QuantifiedMethodology,WilsonIII2014QuantifiedDissipation, Goodrich2018MMSCrossing, Chen2018ElectronShock}, the details of this energy conversion and output remain unclear. The kinetic-scale processes that perform this energy conversion are not well known or well observed within the terrestrial bow shock. Moreover, it is not clear what the resulting energy budget is once the plasma traverses the shock or how it varies for different shock conditions. In this section, we describe the scientific motivation and we describe the measurements necessary to address these questions. \\
\\
\textit{1) What is the partition of energy across collisionless shocks?}

\begin{wrapfigure}{r}{0.5\textwidth}
  \centering
  \vspace{-10pt}
    \includegraphics[width=0.45\textwidth]{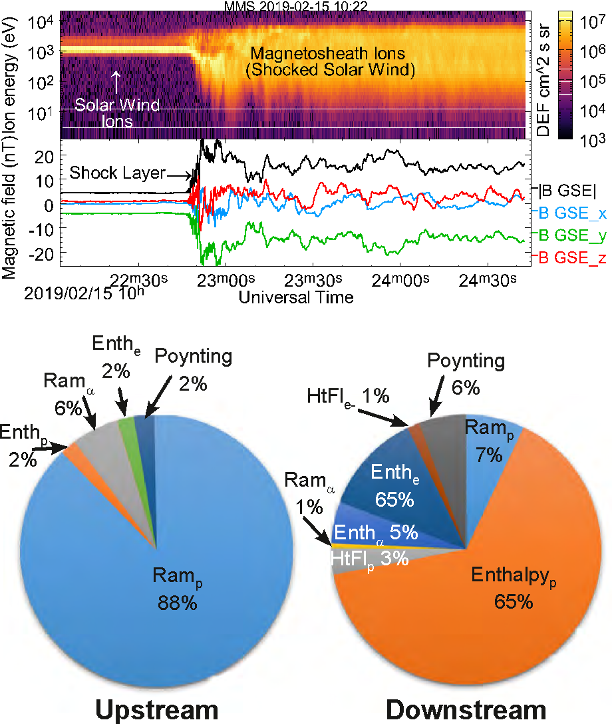}
  \caption{Documented energy partition upstream and downstream of an example shock. (after \cite{Schwartz2022EnergyShocks}). The upstream energy flux is dominated by the proton bulk flow ram energy. It gets partitioned across the particle populations downstream. The relative proportions depend on the upstream parameters in unknown ways.}\label{fig:ener}
\end{wrapfigure}

In order to understand how energy is partitioned in the shock, it is important to accurately resolve the types and weights of different energy inputs and outputs of the system. Simultaneously relating upstream and downstream conditions remains a persistent challenge in studying shock physics as it necessitates simultaneous, complementary, and inter-calibrated upstream and downstream measurements of the plasma.

Important energy fluxes include those related to particle bulk flow, thermal and energetic/nonthermal energy for multiple, relevant species (protons, alphas, heavy ions, and electrons), together with electromagnetic energy. The different energies related to particle species require  measurements of full velocity distribution functions. The thermal properties, anisotropies, and non-Maxwellian thermal features of the cool incident solar wind populations (i.e., electrons, protons, and alphas < 1 keV) as well as higher energy particles (i.e., electrons, H, He, C, N, O, Ne, and Fe > 1 keV) must also be resolved. Crucially, the cold thermal solar wind plasma beam must be fully resolved without compromising the measurement of the hot, shocked plasma or suprathermal reflected and accelerated particles. 

The majority of the upstream energy flux consists of proton ram energy flux while proton enthalpy flux comprises the majority of the downstream partition \cite{Schwartz2022EnergyShocks}, as seen in Figure \ref{fig:ener}. The shock can also produce  other significant energy fluxes including that in accelerated particles, nonthermal features and DC/AC Poynting flux or turbulence. Although these energy fluxes are considered minor contributions to the energy partition, they can be significant to the overall dynamics of the shock, or to the nature of the interaction of the shocked plasma with the magnetosphere. 

Two crucial factors must be considered here. Firstly, the upstream and downstream plasma must be observed in correlation in order to ensure that the output energy fluxes are matched to the measured inputs. Secondly, the  upstream plasma must be measured in such a way that it is clearly not perturbed by conditions of the shock itself, i.e. reflected particles, ultra-low frequency waves, and foreshock phenomena. 

Historically, magnetospheric missions have lacked one or more capabilities to solve this problem. Those capabilities include matched up/downstream measurements, comprehensive inter-calibrated instrumentation, time resolution, velocity-space resolution, and spacecraft separations. Progress can be made with observations from MMS (as shown in Figure \ref{fig:ener}), but significant uncertainties plague our ability to find closure. These uncertainties are fully detailed in \cite{Schwartz2022EnergyShocks}, and partially summarised in the following section.

Without a full account of the energy partition, our modeling and simulation knowledge of shocks, and the applicability of that knowledge to more distant space environments, is at a significant disadvantage. Improvements can and must be made to allow for these observations. MAKOS will do this by engaging four spacecraft with varied spacing. Two of the four spacecraft (separated at ion kinetic scales, $\sim$1000 km)  will act as upstream monitors with apogees up to 25 Earth radii. The two remaining spacecraft  (also spaced $\sim$1000 km apart) will be separated from the upstream monitors by several Earth radii anti-sunward in order observe the resulting magnetosheath. More details are described in our corresponding MAKOS mission white paper.\\
\\
\textit{2) What are the processes governing energy conversion at and within collisionless shocks?}
The knowledge of several different conversion mechanisms have been listed. These include, but are not limited to, a cross-shock electrostatic potential \cite{Chen2018ElectronShock, Tsurutani1981UpstreamResults}, current-driven instabilities such as the Buneman \cite{Goodrich2018MMSCrossing, Bale2007MeasurementShock} and electron-cyclotron drift instability \cite{Breneman2013STEREOMagnetosheath}, magnetic reconnection \cite{Wang2016IonMMS, Gingell2017MMSShock}, and other wave-particle interactions \cite{Chen2018ElectronShock, Goodrich2019ImpulsivelyShocks, Vasko2018SolitaryShocks}, and particle acceleration and reflection. We know, for example, that at even modest Mach number $Q_\perp$ shocks, particle reflection initiates the dispersal in velocity space that results in a higher second moment (temperature). The balance between that mechanism and others within the shock layer that act on both the incident protons and other species is not understood.   It is also unknown how these mechanisms change with upstream conditions, or if the presence of one mechanism drastically alters the resultant downstream plasma.

Within the shock,  energy is converted on the kinetic scale (see above references). This inherently renders MHD modeling insufficient to accurately simulate collisionless shocks in their full complexity. We have learned much from PIC and Vlasov simulations, but we have yet to provide observational confirmation. Historically, in-situ spacecraft have relied on particle detectors that can resolve full velocity distribution functions (VDFs) over one full spin period, on the order from one to tens of seconds. Observed bow shock crossings can have observational lifetimes on the order of seconds, rendering past particle resolution insufficient.

\begin{wrapfigure}{l}{0.6\textwidth}
   \centering
    \vspace{-15pt}
    \includegraphics[width=0.58\textwidth]{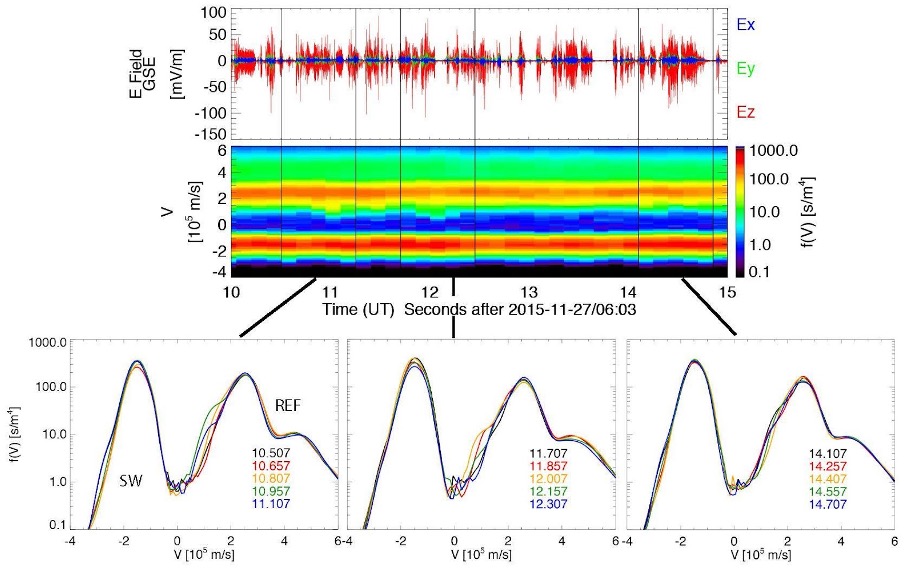}
  \caption{Comparison of 2D ion velocity distributions and electrostatic waves. Bursty electric fields (top) can be linked to fast time variations in particle reflection off the shock which manifests itself in fine scale structure in velocity. This illustrates the interplay between the macroscopic shock inputs and the processes responsible for ultimately converting that energy to other forms. (Taken from \cite{Goodrich2019ImpulsivelyShocks})}\label{fig:waves}
\end{wrapfigure}

MMS shock observations, with high-temporal resolution, allow us to correlate wave and particle behavior like never before (see Figure \ref{fig:waves}), Despite its capabilities, however, MMS has significant limitations in its capability to observe shock phenomena.  We describe these limitations in detail within the following section.

To bring closure to this question, we must measure full velocity distribution functions at a high time resolution (10s of ms) with an energy and angular resolution specified for the solar wind ion distribution. The proposed MAKOS mission intends to develop and outfit such particle instruments. In addition to the DC fields that govern the lowest order particle dynamics, MAKOS will also measure high frequency electric and magnetic field oscillations to identify local plasma instabilities and estimate the amount of energy carried away from the shock region by plasma waves. Using these measurements, plasma instabilities and energy conversion mechanisms will be quantified and distinguished within the shock and then correlated with the energy budgets measured by the spacecraft situated upstream and downstream of the shock. \\
\\
\textit{3) How and why do these processes vary with macroscopic shock parameters?}

The final question is how the energy partitioning process and outputs are related to the shock’s driving conditions. It is known that $\theta_{Bn}$ can influence the geometry and size of the shock, as well as its deviation from laminar behavior. It is not known, however, how $\theta_{Bn}$ can influence the energy budget or energy conversion processes that may occur. The same can be said of the upstream fast magnetosonic Mach number ($M_f$) and plasma beta ($\beta$), the presence of He$^2+$ and/or other minor ion populations, thermal anisotropies, temperatures of both electrons and the various ion species, and the contributions of energetic particle populations. 

The employment of MAKOS will answer this question by observing a statistically significant number of shock crossings with a range of driving conditions, quantifying parametric dependencies of various energy partitioning configurations and energy conversion processes vs. shock orientations and driving conditions. The dataset that will result from MAKOS will provide measurements of >500 quasiparallel and quasi-perpendicular shocks each, assuming they are each observed with approximately equal probability. This will provide sufficient statistics to identify trends in the energy budget and identified energy conversion processes due to specific shock input conditions. Furthermore, the MAKOS orbits offer year-round coverage in the solar wind, also enabling MAKOS to study interplanetary shocks and further bolster the statistics on various shock driving conditions and behavior.

\section*{Limitations of MMS}
MMS is the most sophisticated technology we currently have to measure space plasma in-situ. It can measure full electron velocity distributions over a 30 ms cadence and partial distributions as low as 7.5 ms. It is the most capable mission we have to observe microscale phenomena in the bow shock. And indeed it has, and opened up a completely new avenue into the discussion of the physics that take place in collisionless shocks. However, MMS cannot provide scientific closure to the stated questions concerning collisionless shocks. In this section, we outline the most critical reasons behind this statement.

\begin{wrapfigure}[21]{r}{0.5\textwidth}
    \vspace{-10pt}
   \centering
    \includegraphics[width=0.48\textwidth]{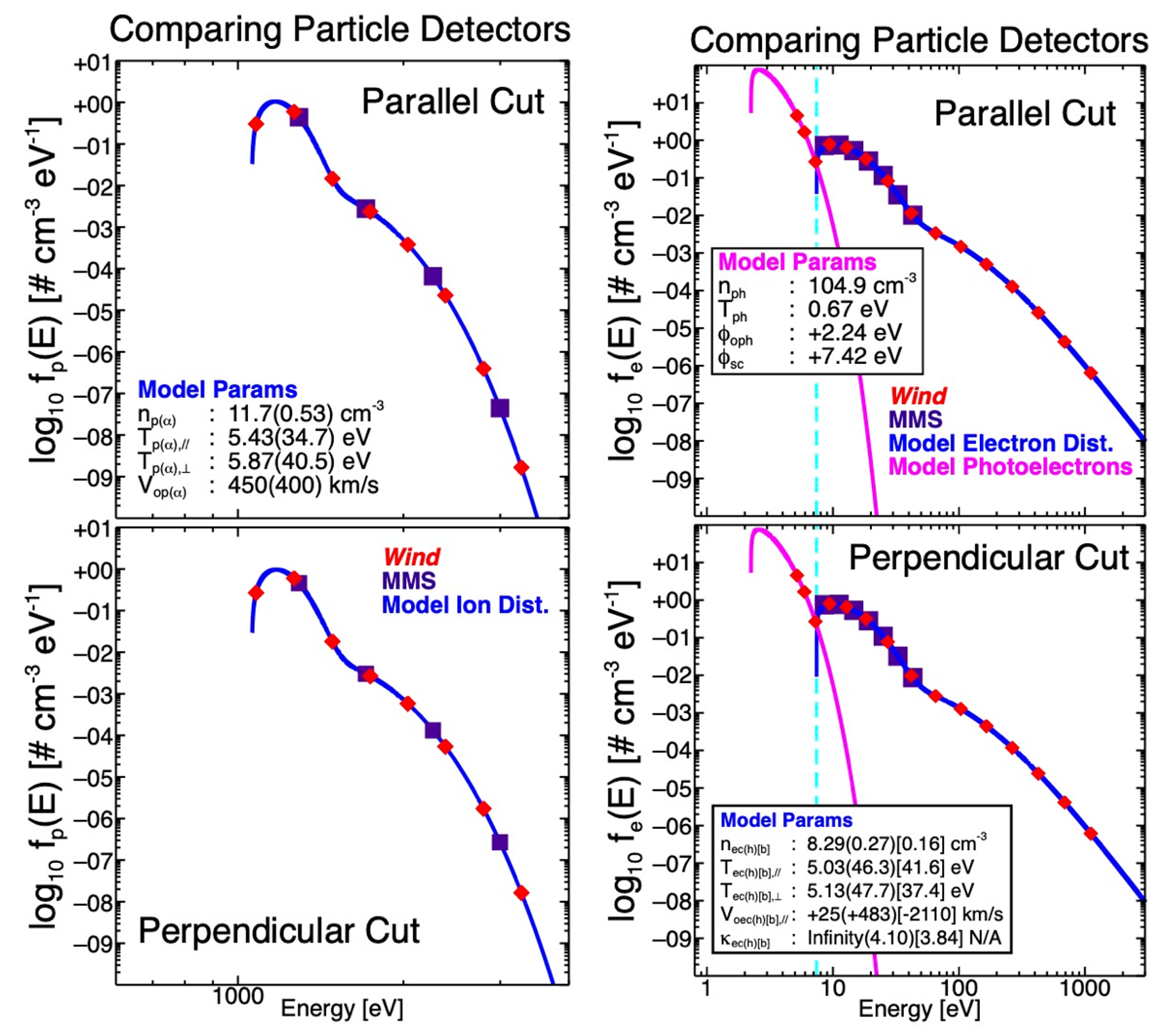}
  \caption{Comparison of MMS and Wind proton (left) and electron (right) energy resolution overlayed on model solar wind VDFs. Note MMS's inability to resolve the solar wind proton peak, and the lack of statistically significant electron counts above $\sim500$\,eV.}\label{fig:dists}

\end{wrapfigure}

Firstly and most critically, MMS cannot resolve the ion solar wind beam distribution. Due to its design, MMS particle detectors (both FPI and HPCA) are not optimized to resolve the proton energy distribution of the solar wind. Figure~\ref{fig:dists} shows the MMS energy coverage of  modeled solar wind populations in comparison to Wind. The proton core population is insufficiently resolved to determine even basic moments such as density and temperature. Nor can the strahl electron population above $\sim500$\,eV be captured as count rates fall below statistical significance. Without accurate resolution of these populations, we cannot characterize the upstream plasma nor observe  the solar wind development through the shock. Please see the white paper written by Wilson et al. for more details.

Secondly, the electric field probes are too long to accurately measure high frequency wave phenomena (see Figure \ref{fig:waves} from \cite{Goodrich2018MMSCrossing}). Observed short wavelength waves appear highly attenuated from the very long boom lengths, rendering them very difficult to analyze. We can resolve this through careful interferometry and application of theory. However, assumptions will always be made to do so and we therefore cannot make significant progress to understanding the roles waves take in energy conversion within the terrestrial bow shock.

Finally, the MMS spacecraft separation distances do not allow for appropriate simultaneous upstream and downstream measurements. MMS has had an average of $\sim$15 km separation in the dayside magnetosphere, well within the solar wind gyroradius ($\sim$1000 km). This is not sufficient distance to determine the conditions of unperturbed solar wind. These scales can be adjusted, and plans are currently implemented to enable cross-scale measurements within the realm of the bow shock and magnetosheath. However, even if appropriate distances can be achieved, the two previous outstanding issues remain.

\section*{Expected Scientific Impact}
The full knowledge of shock micro-processes will more firmly establish our knowledge of fundamental plasma processes. This will further enable collaboration with the laboratory plasma community, as they develop and experiment with similar scale and mechanisms. This will also enable greater collaboration with the astrophysical community, as they observe astrophysical shocks via remote sensing. The observed radiation from these shocks stem from the post-energy conversion process. By acquiring an accurate knowledge of energy partitioning resulting from collisionless shocks, we will establish clearer connections to the processes and implications in shocks beyond our in-situ capabilities.

\section*{Summary and Conclusions}
In all applications of space plasmas, three universal plasma processes dominate the dynamics. Magnetic reconnection reconfigures topologies, allows plasma mixing and can drive flows and acceleration. Turbulence transfers energy to small scales where it can be efficiently dissipated.

Collisionless shocks are a fundamental plasma process. They are the prime ``thermalizers" (converting flow energy to heat) and ``non-thermalizers'' (converting flow energy to nonthermal features and energetic particles) in the astrophysical world. Despite that importance, and decades of observations and thoeretical/simulational studies, the basic ability to predict how a shock with given upstream parameters will partition the incident energy amongst the various degrees of freedom available remains elusive. This white paper has laid out the questions that need to be answered to address this ignorance, the reasons why existing missions and datasets cannot provide a complete answer, and the capabilities a dedicated mission must have in order to do so.

The heliophysics community recognizes the importance of fundamental processes through the support of the previously launched Magnetospheric Multiscale (MMS) mission and the recently selected Mid-Explorer Helioswarm mission. Both were selected with the intent of observing magnetic reconnection and plasma turbulence respectively. In order to achieve a complete view of the fundamental physics that dominate our universe, collisionless shocks must also be considered a subject of importance in heliophysics. This can and must be done by supporting targeted opportunities to observe the terrestrial bow shock in-situ, starting with MAKOS.

\pagebreak
\bibliography{MAKOS_Science}
\end{document}